\documentclass[twocolumn,aps,pra,groupedaddress,showpacs]{revtex4}
\usepackage{epsfig,amssymb}
\def\comment#1{}\def\labell#1{\label{#1}}
\begin{document}
\title{The role of entanglement in dynamical evolution}
\author{Vittorio Giovannetti$^1$, Seth Lloyd$^{1,2}$, and Lorenzo
Maccone$^1$}\affiliation{$^1$Massachusetts Institute of Technology --
Research Laboratory of Electronics\\$^2$Massachusetts Institute of
Technology -- Department of Mechanical Engineering\\ 77
Massachusetts Ave., Cambridge, MA 02139, USA}

\begin{abstract}
Entanglement or entanglement generating interactions permit to achieve
the maximum allowed speed in the dynamical evolution of a composite
system, when the energy resources are distributed among
subsystems. The cases of pre-existing entanglement and of
entanglement-building interactions are separately addressed. The role
of classical correlations is also discussed.
\end{abstract}
\maketitle
The problem of determining how to exploit the available resources to
achieve the highest evolution speed is relevant to deriving physical
limits in a variety of contexts. Of particular interest are, for
example, the maximum rate for information processing
{\cite{margolus,sethlimits}} or for information exchange through
communication channels {\cite{caves,seth}}.

The time-energy uncertainty relation imposes a lower limit on the time
interval ${\cal T}_\perp$ that it takes for a quantum system to evolve
through two orthogonal states {\cite{uncer,margolus,nostro}}.  This bound on
${\cal T}_\perp$ is related to the spread in energy of the system.
Recently, Margolus and Levitin have linked such a quantity also to the
average energy of the system {\cite{margolus}}. These two
conditions together establish the {\it quantum speed limit time}, i.e.
the minimum time ${\cal T}(E,\Delta E)$ required for a system with
energy $E$ and energy spread $\Delta E$ to evolve through two
orthogonal states (see Sec.~\ref{s:qsl}).

In this paper we analyze the ${\cal T}(E,\Delta E)$ of systems
composed of $M$ subsystems, focusing on the role of correlations
between the $M$ components and separately addressing the cases of
non-interacting and interacting subsystems. When no interactions are
present, we first show that for initially-separable pure states, the
quantum speed limit is achievable only in the asymmetric situation in
which only one of the subsystems evolves in time and carries all the
system's energy resources. We then provide an example that shows that
the presence of entanglement in the initial state allows a dynamical
speedup also when the energy resources are homogeneously distributed
among the subsystems. In this way, showing that homogeneous separable
states cannot exhibit speedup while at least one homogeneous entangled
case that exhibits speedup exists, we prove that entanglement is
necessary to achieve the quantum speed limit, at least in the case of
pure states (see Sec.~\ref{s:non}). When classical mixtures are taken
into account, the situation is more complex: energy-homogeneous
separable states that reach the bound do exist. However, their
ensemble realizations are either mixtures of entangled configurations
or mixtures of energy-asymmetric configurations: more precisely, each
separable unraveling of the state
$\varrho=\sum_np_n\varrho_1^{(n)}\otimes\cdots\otimes\varrho_M^{(n)}$
must be composed by product states in which only a single subsystem
evolves rapidly to an orthogonal configuration while the other ones do
not evolve at all (see Sec.~\ref{s:mix}).

In the case of interacting subsystems (see Sec.~\ref{s:int})
homogeneous pure unentangled states can still achieve the   ${\cal
T}(E,\Delta E)$ bound. It will be shown that the reason for this
behavior is the entanglement built up during the interaction.

\section{Quantum speed limit time}\labell{s:qsl}
Consider a system in an initial state $|\Psi\rangle$ of mean energy
$E=\langle\Psi|H|\Psi\rangle$ (where $H$ is the Hamiltonian and where
we assume zero ground state energy). The Margolus-Levitin theorem
{\cite{margolus}} asserts that it takes at least a time ${\cal
T}_\perp\geqslant{\pi\hbar}/({2E})$ for the system to evolve from
$|\Psi\rangle$ to an orthogonal state. This result complements the
time-energy uncertainty relation, which requires ${\cal
T}_\perp\geqslant{\pi\hbar}/({2\Delta E})$, 
where $\Delta
E=\sqrt{\langle\Psi|(H-E)^2|\Psi\rangle}$ is the energy spread of the
state {\cite{uncer,margolus}}. The Margolus-Levitin theorem gives a
better bound on ${\cal T}_\perp$ than the uncertainty relations when
an asymmetric energy distribution yields $\Delta E>E$. Joining the two
above inequalities one obtains the quantum speed limit time, i.e. the
minimum time ${\cal T}(E,\Delta E)$ required for the evolution to an
orthogonal state, as
\begin{eqnarray}
{\cal T}_\perp\geqslant{\cal
T}(E,\Delta E)\equiv\max\left(\frac{\pi\hbar}{2E}\;,\
\frac{\pi\hbar}{2\Delta E}\right)
\;\labell{qsl}.
\end{eqnarray}
In {\cite{margolus}} it has been shown that states that saturate this
bound do exist. In the appendix the bound (\ref{qsl}), derived
only for pure states in {\cite{margolus}}, is shown to apply also for
mixed states.

In this paper we analyze the quantum speed limit time (\ref{qsl}) of
systems composed of $M$ parts. The Hamiltonian is of the form
\begin{eqnarray}
H=\sum_{k=1}^MH_k+H_{{\rm int}}
\;\labell{hamilt1},
\end{eqnarray}
where the $H_k$ are the free Hamiltonians of the subsystems and
$H_{{\rm int}}$ is a non-trivial interaction Hamiltonian between them.
When $H_{\rm int}=0$, the Hamiltonian is not able to generate
correlations between the subsystems so that they evolve independently.
This case is analyzed in the following section, where it is shown
that, unless correlations are present in the initial state of the
system, the energetic resources available cannot be efficiently used
to achieve the bound (\ref{qsl}) when they are distributed among the
$M$ parts.

\section{Non interacting subsystems}\labell{s:non}
In this section we show that, for non-interacting subsystems, pure
separable states cannot reach the quantum speed limit unless all
energy resources are devoted to one of the subsystems. This is no more
true if the initial state is entangled: a cooperative behavior is
induced such that the single subsystems cannot be regarded as
independent entities.

A separable pure state has the form
\begin{eqnarray}
|\Psi_{sep}\rangle=|\psi_1\rangle_1\otimes\cdots\otimes|\psi_M\rangle_M
\;\labell{stcl},
\end{eqnarray}
where $|\psi_k\rangle_k$ is the state of the $k$-th subsystem which
has energy ${\cal E}_k$ and spread $\Delta{\cal E}_k$.  Since there is
no interaction ($H_{\rm int}=0$), the vector $|\Psi_{sep}\rangle$
remains factorizable at all times. It becomes orthogonal to its
initial configuration if at least one of the subsystems evolves to an
orthogonal state. The time employed by this process is limited by the
energy and the energy spread of each subsystem, through
Eq.~(\ref{qsl}). By choosing the time corresponding to the ``fastest''
subsystem, the time ${\cal T}_\perp$ for the state
$|\Psi_{sep}\rangle$ is {\cite{nota1}}
\begin{eqnarray} {\cal
T}_\perp\geqslant\max\left(\frac{\pi\hbar}{2{\cal E}_{max}}\;,\
\frac{\pi\hbar}{2\Delta {\cal E}_{max}}\right)
\;\labell{qsl1},
\end{eqnarray}
where ${\cal E}_{max}$ and $\Delta{\cal E}_{max}$ are the maximum
values of the energy and energy spread of the $M$ subsystems.  For
the state $|\Psi_{sep}\rangle$, the total energy is $E=\sum_k{\cal
  E}_k$ and the total energy spread is $\Delta
E=\sqrt{\sum_k\Delta{\cal E}^2_k}$. This implies that the bound
imposed by Eq.~(\ref{qsl1}) is always greater or equal than ${\cal
  T}(E,\Delta E)$ of Eq.~(\ref{qsl}), being equal only when ${\cal
  E}_{max}=E$ or $\Delta{\cal E}_{max}=\Delta E$, e.g. when one of the
subsystems has all the energy or all the energy spread of the whole
system. This means that only such subsystem is evolving in time: the
remaining $M-1$ are stationary. The gap between the bound (\ref{qsl1})
for separable pure states and the bound (\ref{qsl}) for arbitrary
states reaches its maximum value for systems that are homogeneous in
the energy distribution, i.e. such that ${\cal E}_{max}=E/M$ and
$\Delta{\cal E}_{max}=\Delta E/\sqrt{M}$. In this case,
Eq.~(\ref{qsl1}) implies that, for factorizable states, one has at
least ${\cal T}_\perp\geqslant\sqrt{M}\;{\cal T}(E,\Delta E)$. In
fact, if $E\geqslant\Delta E$, ${\cal T}_\perp$ is always greater than
$M$ times the quantum speed limit time. On the other hand, if $\Delta
E\geqslant E$, we find that \begin{eqnarray} {\cal
    T}_\perp\geqslant\left\{\matrix{\displaystyle
\sqrt{M}\;{\cal T}(E,\Delta E)
      &&\mbox{for }M\leqslant M^*\;\;\cr \cr\displaystyle\frac{M}{\sqrt{M^*}}
      \;{\cal T}(E,\Delta E)&&\mbox{for }M\geqslant M^*\;,}
    \right.  \;\labell{tp}
\end{eqnarray}
where $M^*\equiv(E/\Delta E)^2$.

In order to show that the bound {\it is} indeed achievable when
entanglement is present consider the following entangled state
\begin{eqnarray} |\Psi_{ent}\rangle=\frac
1{\sqrt{N}}\sum_{n=0}^{N-1}|n\rangle_1\otimes\cdots\otimes|n\rangle_M
\;\labell{stent},
\end{eqnarray}
where $|n\rangle_k$ is the energy eigenstate (of energy
$n\hbar\omega_{0}$) of the $k$-th subsystem. The state
$|\Psi_{ent}\rangle$ is homogeneous since each of the $M$ subsystems
has energy ${\cal E}=\hbar\omega_{0} (N-1)/2$ and $\Delta {\cal
E}=\hbar\omega_{0}\sqrt{N^2-1}/(2\sqrt{3})$. The total energy and
energy spread are given by $E=M{\cal E}$ and $\Delta E=M\Delta {\cal
E}$ respectively.  The scalar product of $|\Psi_{ent}\rangle$ with its
time evolved $|\Psi_{ent}(t)\rangle$ is
\begin{eqnarray} \langle\Psi_{ent}|\Psi_{ent}(t)\rangle=\frac
1N\sum_{n=0}^{N-1}e^{-inM\omega_{0} t}
\;\labell{evoluz},
\end{eqnarray}
where the factor $M$ in the exponential is a peculiar signature of the
energy entanglement.  The value of ${\cal T}_\perp$ for the state
$|\Psi_{ent}\rangle$ is given by the smallest time $t\geqslant 0 $ for
which this quantity is zero, i.e. $2\pi/(NM\omega_{0})$. It is smaller
by a factor $\sim\sqrt{M}$ than what it would be for homogeneous
separable pure states with the same value of $E$ and $\Delta E$, as
can be checked through Eq.~(\ref{qsl1}). The above example can be
easily extended to the general case in which the $H_k$ are not
necessarily identical. The effect shown here can be exploited where it
is necessary to increase the speed of systems while equally sharing
the energy resources among the subsystems. States of the type
$|\Psi_{ent}\rangle$ have been used in {\cite{qps}} in order to
increase the time resolution of traveling pulses.

In summary, pure separable states can reach the quantum speed limit
only in the case of highly asymmetric configurations where one of the
subsystems evolves to an orthogonal configuration at the maximum speed
allowed by its energetic resources, while the other subsystems do not
evolve. In all other cases entanglement is necessary to achieve the
bound. This, of course, does not imply that all entangled states
evolve faster than their unentangled counterparts.

\subsection{Classical mixtures.}\labell{s:mix}
What happens when classical correlations among subsystems are
considered?  A separable state of an $M$-parts composite system can be
always expressed by the following convex convolution   \begin{eqnarray}
\varrho=\sum_np_n\varrho_1^{(n)}\otimes\cdots\otimes\varrho_M^{(n)}
\;\labell{conv},
\end{eqnarray}
where $p_n$ are positive coefficients which sum up to one and where
the normalized density matrix $\varrho_k^{(n)}$ describes a state of
the $k$-th subsystem with energy $E_k^{(n)}$ and energy spread $\Delta
E_k^{(n)}$.  Equation (\ref{conv}) is a mixture of independent product
state configurations, labeled by the parameter $n$, which occur with
probability $p_{n}$: it displays classical correlations but no
entanglement between the $M$ subsystems.  For non-interacting systems,
the energy $E$ and spread in energy $\Delta E$ of $\varrho$ are given
by
\begin{eqnarray}
E&=&\sum_np_n\sum_{k=1}^ME_k^{(n)} \; \labell{energia} \\ \nonumber
\Delta E^2&=&\sum_np_n\left[\sum_{k=1}^M(\Delta E_k^{(n)})^2+
\left(\sum_{k=1}^ME_k^{(n)}-E\right)^2\right] \; ,
\end{eqnarray}
and the state $\varrho(t)$ at time $t$ is always of the form
(\ref{conv}) where $\varrho_k^{(n)}$ are replaced by their evolved
$\varrho_k^{(n)}(t)$.  If $\varrho$ reaches the quantum speed limit
then it is orthogonal to its evolved at time ${\cal T}(E,\Delta E)$,
i.e.
\begin{eqnarray}
\sum_{nm}p_np_m\chi_1^{(n,m)}(t)
\cdots\chi_M^{(n,m)}(t)\Bigr|_{t={\cal T}(E,\Delta E)}=0
\;\labell{ort2}
\end{eqnarray}
where
$\chi_k^{(n,m)}(t)\equiv\mbox{Tr}[\varrho_k^{(n)}(t)\;\varrho_k^{(m)}]$.
Using the spectral decomposition, one immediately sees that all the
terms $\chi$ are non-negative real quantities. This means that
Eq.~(\ref{ort2}) is satisfied if and only if each of the summed terms
is equal to zero independently on $n$ and $m$. In particular, focusing
on the case $n=m$, at least one subsystem must exist (say the
$k_{n}$-th) for which $\chi_{k_{n}}^{(n,n)}[{\cal T}(E,\Delta E)]=0$.
Applying the quantum speed limit (\ref{qsl}) to the state
$\varrho_{k_{n}}^{(n)}$ of this subsystem the following inequality
results
\begin{eqnarray}
{\cal T}(E,\Delta E)\geqslant {\cal T}(E_{k_{n}}^{(n)},\Delta
E_{k_{n}}^{(n)})
\;\labell{gatto}.
\end{eqnarray}
Suppose now that $E \geqslant \Delta E$, i.e. ${\cal T}(E,\Delta E)=
\pi\hbar/(2\Delta E)$.  In this case, from Eqs.~(\ref{energia}) and
(\ref{gatto}) one can show that for each $n$, $\Delta E_{k}^{(n)}=0$
for all ${k}\neq {k_{n}}$, while $ E_{k_{n}}^{(n)}\geqslant \Delta
E_{k_{n}}^{(n)} = \Delta E$.  On the other hand, if $ \Delta
E\geqslant E$ then one can analogously obtain that for each $n$,
$E_{k}^{(n)}=0$ for all ${k}\neq{k_{n}}$, while $ \Delta
E_{k_{n}}^{(n)}\geqslant E_{k_{n}}^{(n)} = E$. In both cases the
inequality (\ref{gatto}) becomes an identity, i.e. the quantum speed
limit time ${\cal T}(E_{k_{n}}^{(n)},\Delta E_{k_{n}}^{(n)})$ of the
state $\varrho_{k_{n}}^{(n)}$ coincides with the quantum speed limit
time ${\cal T}(E,\Delta E)$ of the global state $\varrho$. Moreover,
for all $k\neq k_{n}$ the states $\varrho_{k}^{(n)}$ are eigenstates
of the Hamiltonians $H_{k}$ {\cite{nota2}, i.e. they cannot evolve to
  orthogonal configurations.

This proves that the separable state (\ref{conv}) achieves the quantum
speed limit only if, in any statistical realization $n$ of the system,
a single subsystem evolves to an orthogonal configuration at its own
maximum speed limit time (which coincides with ${\cal T}(E,\Delta E)$
of the whole system). All the other subsystems do not evolve.

Classical correlations among subsystems, however, can produce
configurations $\varrho$ that achieve the speed limit and are
statistically energy-homogeneous: i.e.  {\it in average} all
subsystems share the same resources. As an example, consider the
separable state
$\varrho_{s}=(\varrho_a\otimes\varrho_b+\varrho_b\otimes\varrho_a)/2$
of a bipartite system composed by two identical subsystems (e.g. two
spins), where $\varrho_b$ is the zero energy ground state and
$\varrho_a$ is a normalized density matrix which saturates its own
quantum speed limit. Since the energy $E$ and the energy spread
$\Delta E$ of the $\varrho_{s}$ coincide with those of $\varrho_a$,
these two matrices have the same value of ${\cal T}(E,\Delta E)$.  The
density operator $\varrho_{s}$ describes a mixture where half of the
times the first spin is in the state $\varrho_a$ and the second spin
is in the ground state, and in the other half their roles are
exchanged: of course in this configuration in average the two spins
are in the same state $(\varrho_a+\varrho_b)/2$.  Assume now that
$\mbox{Tr}[\varrho_a(t)\;\varrho_b]=\mbox{Tr}[\varrho_b(t)\;\varrho_a]=0$:
in this case $\varrho_{s}$ will saturate the quantum speed limit.

Since all the above derivation applies for separable unravellings of
the form (\ref{conv}), one can say that in each experimental run only
one of the subsystems evolves to an orthogonal state. Of course the
state $\varrho$ allows also unravellings that are not of the form
(\ref{conv}) in which the statistical realizations may contain
entanglement between subsystems (e.g. a fully mixed state can be
obtained from a statistical mixture of maximally entangled states).
The above derivation does not apply to these entangled decompositions
of $\varrho$, yet the role of entanglement is self-evident in this
case. Hence, practically, there are two different ways to build
``fast'' separable states through classical correlations: either
starting from the separable configurations (\ref{conv}) in which only
one of the subsystems evolves, or starting from entangled
configurations. What is definitely impossible is to build a $\varrho$
that reaches the bound by mixing separable configurations in which the
energy is not concentrated in one of the subsystems.

\section{Interacting subsystems}\labell{s:int}
For the sake of simplicity, in analyzing interacting subsystems, we
focus only on the pure state case where the effects of entanglement
are more evident. Two situations are possible.  Either $H_{\rm int}$
does not introduce any entanglement in the initial state of the system
or $H_{\rm int}$ builds up entanglement among subsystems. In the first
case, no correlations among the subsystems are created so that each
subsystem evolves independently as
$|\Psi(t)\rangle=\otimes_{j=1}^M|\psi_j(t)\rangle_j$, unless
entanglement was present initially. Since this type of evolution can
always be described as determined by an interaction-free effective
Hamiltonian, the results of the previous section apply.  In the second
case, when $H_{\rm int}$ builds up entanglement, the system may reach
the bound even though no entanglement was already present initially.
In fact, as will be shown through an example, one can tailor suitable
entangling Hamiltonians that speed up the dynamical evolution even for
initial homogeneous separable states.

An interaction capable of speeding up the dynamics is given by the
following Hamiltonian for $M$ qubits
\begin{eqnarray}
H=\hbar\omega_0\sum_{k=1}^M(1-\sigma_x^{(k)})+\hbar\omega(1-S)
\;\labell{hamilt},
\end{eqnarray}
where $\sigma_x^{(k)}$ is the Pauli operator $|1\rangle\langle
0|+|0\rangle\langle 1|$ for the $k$-th qubit and where
$S=\prod_{k=1}^M\sigma_x^{(k)}$. The first term in Eq.~(\ref{hamilt})
is the free Hamiltonian which rotates independently each of the qubits
at frequency $\omega_0$.  The second term is a global interaction
which rotates collectively all the qubits at frequency $\omega$,
coupling them together. Consider an initial factorized state where all
qubits are in eigenstates of the $\sigma_z^{(k)}$ Pauli matrices,
i.e.\begin{eqnarray}
|\Psi\rangle\equiv|J_1\rangle_1\otimes\cdots\otimes|J_M\rangle_M
\;\labell{statoiniz},
\end{eqnarray}
where $J_k$ are either $0$ or $1$. This is an homogeneous
configuration of the system. Moreover, the energy $E=\hbar(\omega +
M\omega_0)$ and the energy spread   $\Delta E=\hbar\sqrt{\omega^2 +
M\omega_0^2}$ of this state give ${\cal T}(E,\Delta
E)=\pi/(2\sqrt{\omega^2 + M\omega_0^2})$. The state $|\Psi\rangle$
evolves to the entangled configuration
\begin{eqnarray} 
|\Psi(t)\rangle=e^{-iEt/\hbar}\left[
\cos(\omega t)|J(t)\rangle
+i\sin(\omega t)|\overline{J(t)}\rangle\right]
\;\labell{statot},
\end{eqnarray}
where \begin{eqnarray}
|J(t)\rangle=\bigotimes_{k=1}^M\left[
\cos(\omega_0 t)|J_k\rangle_k
+i\sin(\omega_0 t)|\overline{J_k}\rangle_k\right]
\;\labell{jaydit},
\end{eqnarray}
with the overbar denoting qubit negation
($|\overline{0}\rangle\equiv|1\rangle$,
$|\overline{1}\rangle\equiv|0\rangle$). Imposing the orthogonality
between $|\Psi\rangle$ and $|\Psi(t)\rangle$, we find that ${\cal
T}_\perp$ is the minimum value of $t$ for which \begin{eqnarray}
\cos(\omega t)\cos^M(\omega_0 t)+i^{M+1}\sin(\omega t)\sin^M(\omega_0
t)=0 
\;\labell{condiz}.
\end{eqnarray}
From Eq.~(\ref{condiz}) it is easy to check that, for $\omega=0$ (no
interaction) ${\cal T}_\perp$ is $\sqrt{M}$ times bigger than   ${\cal
T}(E,\Delta E)$. Instead, for $\omega_0=0$ (no free evolution) the
system reaches the speed limit, i.e.    ${\cal T}_\perp={\cal
T}(E,\Delta E)$. In Fig.~\ref{f:grafico} the value of ${\cal T}_\perp$
is compared to the value of ${\cal T}(E,\Delta E)$ for different
values of $\omega$, showing that as the interaction becomes
predominant, ${\cal T}_\perp$ tends to ${\cal T}(E,\Delta E)$. 

This example shows that a suitable $H_{\rm int}$ can allow a
homogeneous pure state to reach the quantum speed limit. In order to
reach this bound, however, the interaction must {\it i)} connect all
the qubits and {\it ii)} be sufficiently strong (see
Fig.~\ref{f:grafico}). A simple counterexample for {\it i)} can be
obtained by considering the case in which the $M$ qubits are divided
in $G$ non-interacting groups which have an Hamiltonian of the same
form of (\ref{hamilt}) and contain $Q=M/G$ qubits each. In this case
entanglement cannot build up between qubits of different groups and it
is immediate to see that ${\cal T}_\perp$ is at least
$\sqrt{M/Q}\;{\cal T}(E,\Delta E)$. The order $K$ of the interaction
(i.e. the number of subsystems that are involved in a single vertex of
interaction) also plays an important role: the example illustrated by
the Hamiltonian (\ref{hamilt}) describes an $M$-th order case. For any
given $K$, a rich variety of cases are possible depending on how the
interaction is capable of constructing entanglement between
subsystems.  The simplest example is an Ising-like model where there
is a $K$-th order coupling between neighbors in a chain of qubits.
Here one only has a $\sqrt{K}$ improvement over the non-interacting
case {\cite{spiepaper}}.

\begin{figure}[hbt]
\begin{center}
\epsfxsize=.7
\hsize\leavevmode\epsffile{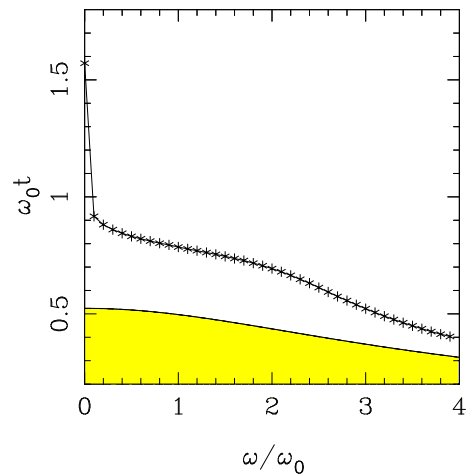}
\end{center}
\caption{Plot of ${\cal T}_\perp$ from Eq.~(\ref{condiz}) (asterisks)
and of ${\cal T}(E,\Delta E)$ of the state (\ref{statoiniz}) (dashed
line) as a function of the relative intensity of the interaction
Hamiltonian $\omega/\omega_0$ for $M=9$. The lower shaded region is
the area forbidden by Eq.~(\ref{qsl}). }
\labell{f:grafico}\end{figure}

A recent proposal {\cite{seth}} uses the effect described in this
section to increase the communication rate by a factor $\sqrt{M}$ over
a communication channel composed of $M$ independent parallel channels
which uses the same resources.

\section{Conclusions}\labell{s:concl}
In conclusion we have studied the quantum speed limit for composite
systems. We have analyzed the role of correlations (quantum and
classical) among subsystems emphasizing the role of entanglement. The
Hamiltonians that do not create quantum correlations need to operate
on initially entangled states in order to speed up the dynamics
(except for the special case in which only one subsystem evolves). On
the other hand, entanglement-generating Hamiltonians are capable of
speeding up the dynamics even starting from separable configurations.

\appendix
\section{The quantum speed limit time for mixed states }\labell{s:mml}
Here the quantum speed limit, which was proved for pure states in
{\cite{margolus}}, is extended to mixed states.  The quantity   ${\cal
T}_\perp$ is defined as the minimum time $t$ for which the evolved
density matrix $\varrho(t)$ of a system of energy $E$ and energy
spread $\Delta E$ becomes orthogonal to the initial state $\varrho$,
i.e.  $\mbox{Tr}[\varrho(t)\varrho]=0$. Using the spectral
decomposition, $\varrho$ can be written as
$\varrho=\sum_{n}\lambda_n|\phi_n\rangle\langle\phi_n|$, where
$\lambda_n$ are positive coefficients which sum up to one and where
$\{|\phi_n\rangle\}$ is an orthonormal set.  From the definition of
${\cal T}_\perp$ it then follows that
\begin{eqnarray} 
{\cal T}_\perp\geqslant
&=&\min_{t\geqslant 0}\Big(\langle
\phi_n(t)|
\phi_m\rangle=0\quad \forall n,m\Big)\nonumber\\
&\geqslant&
\min_{t\geqslant 0}\Big(\langle
\phi_n(t)|
\phi_n\rangle=0\Big)\quad\forall n
\;\labell{tperp1},
\end{eqnarray}
where $|\phi_n(t)\rangle$ is the time-evolved of $|\phi_n\rangle$.
Applying Eq.~(\ref{qsl}) to the pure states $|\phi_n\rangle$, one
finds
\begin{eqnarray} {\cal T}_\perp\geqslant\max
\left(\frac{\pi\hbar}{2E_{min}},\frac{\pi\hbar}{2\Delta
E_{min}}\right)
\;\labell{tperp2},
\end{eqnarray}
where $E_{min}$ and $\Delta E_{min}$ are respectively the minima on
$n$ of the energy $E_n$ and of the spread $\Delta E_n$ of the state
$|\phi_n\rangle$. Since for the state $\varrho$ the energy and the
energy spread are such that
\begin{eqnarray}
E&=&\sum_n\lambda_nE_n
\geqslant E_{min}\labell{as}\\\nonumber\Delta E&=&\sqrt{\sum_n\lambda_n[\Delta
E_n^2+(E-E_n)^2]}\geqslant\Delta E_{min}\;,
\end{eqnarray}
 from Eq.~(\ref{tperp2}) the
quantum speed limit follows.

This work was funded by the ARDA, NRO, NSF, and by ARO under a MURI
program.


\begin{references}
\bibitem{nota1}{If $H=\sum_{k}H_{k}$ has zero energy
ground eigenstate, one can redefine all $H_{k}$ to have zero minimum
eigenvalue.}
\bibitem{nota2}{If $\Delta E\geqslant E$ this comes from the fact that
    for $k\neq k_n$, all the $\varrho_k^{(n)}$ have zero mean energy
    and $H_{k}$ have zero energy ground state.}
\end{references}
\end{document}